\documentclass[12pt]{iopart}

\usepackage{iopams}
\usepackage{epsfig}
\begin{document}

\title{ Nuclear Suppression of Jets and $R_{AA}$  at the LHC}

\author{G.Y.~Qin, J.~Ruppert, S.~Turbide, C.~Gale and S.~Jeon}

\address{Department of Physics, McGill University, Montreal, QC, H3A 2T8, Canada}

\begin{abstract}
The nuclear modification factor $R_{AA}$ for charged hadron production at the LHC is predicted from jet energy
loss induced by gluon bremsstrahlung. The Arnold, Moore, and Yaffe \cite{Arnold:2001ms}  formalism is used,
together with an ideal hydrodynamical model \cite{Eskola:2005ue}.

\end{abstract}


We present a calculation of the nuclear modification factor $R_{AA}$ for charged hadron production as a function
of $p_T$ in Pb+Pb collisions at $\sqrt{s}=5.5~{\rm ATeV}$ in central collisions at mid-rapidity at the LHC. The
net-energy loss of the partonic jets is calculated by applying the Arnold, Moore, and Yaffe (AMY) formalism to
calculate gluon bremsstrahlung \cite{Arnold:2001ms}. The details of jet suppression relies on an understanding
of the nuclear medium, namely the temperatures and flow profiles that are experienced by partonic jets while
they interact with partonic matter at $T \geq T_c$. Our predictions use a boost-invariant ideal hydrodynamic
model with initial conditions calculated from perturbative QCD + saturation \cite{Eskola:2005ue,Eskola2}. It is
emphasized that the reliability of this work hinges on the validity of hydrodynamics at the LHC.  It has been
verified that $R_{AA}$ for $\pi_0$ production as a function of $p_T$ as obtained in the same boost-invariant
ideal hydrodynamical model adjusted to Au+Au collisions at $\sqrt{s}=0.2~{\rm ATeV}$ \cite{Eskola:2005ue} is in
agreement with preliminary data from PHENIX in central collisions at RHIC (and the result is very close to the
one obtained in 3D hydrodynamics presented in \cite{Qin:2007ys}). In AMY the strong coupling constant $\alpha_s$
is a direct measure of the interaction strength between the jet and the thermalized soft medium and is the only
quantity not uniquely determined in the model, once the temperature and flow evolution is fixed by the initial
conditions and subsequent hydrodynamical expansion. We found that assuming a constant $\alpha_s=0.33$ describes
the experimental data in most central collisions at RHIC. It is conjectured that $\alpha_s$ should not be
changed very much at the LHC since the initial temperature is about twice larger than the one at RHIC
whereas $\alpha_s$ is only logarithmically dependent on temperature. We present results for $\alpha_s=0.33~{\rm
and}~0.25$.

For details of the calculation of nuclear suppression, we refer the reader to \cite{Qin:2007ys}. The extension
to the LHC once the medium evolution and $\alpha_s$ are fixed is straightforward. The initial jets are produced
with an initial momentum distribution of jets computed from pQCD in the factorization formalism including
nuclear shadowing effects. The probability density $\mathcal{P}_{AB}(\vec{r}\bot)$ of finding a hard jet at the
transverse position $\vec{r}_\bot$ in central A+B collisions is given by the normalized product of the nuclear
thickness functions, $\mathcal{P}_{AB}(\vec{r}_\bot) = {T_A(\vec{r}_\bot)T_B(\vec{r}_\bot})/{T_{AB}}$ and is
calculated for Pb+Pb collisions. The evolution of the jet momentum distribution $P_j(p,t)={dN_j(p,t)}/{dpdy}$ in
the medium is calculated by solving a set of coupled rate equations with the following generic form,
\begin{eqnarray}
\frac{dP_j(p,t)}{dt} = \sum_{ab} \int dk \left[P_a(p+k,t) \frac{d\Gamma^{a}_{jb}(p+k,p)}{dk dt} -
P_j(p,t)\frac{d\Gamma^{j}_{ab}(p,k)}{dk dt}\right],
\end{eqnarray}
where ${d\Gamma^{j}_{ab}(p,k)}/{dk dt}$ is the transition rate for the partonic process $j\to a+b$ which depends
on the temperature and flow profiles experienced by the jets traversing the medium. The hadron spectrum
${dN^h_{AB}}/{d^2p_Tdy}$ is obtained by the fragmentation of jets after their passing through the medium. The
nuclear modification factor $R_{AA}$ is  computed as
\begin{eqnarray}
R^h_{ AA}(\vec{p}_T,y) = \frac{1}{N_{coll}} \frac{{dN^h_{ AA}}/{d^2p_Tdy}} {{dN^h_{ pp}}/{d^2p_Tdy}}.
\end{eqnarray}

\begin{figure}[htb]
\begin{center}
\epsfig{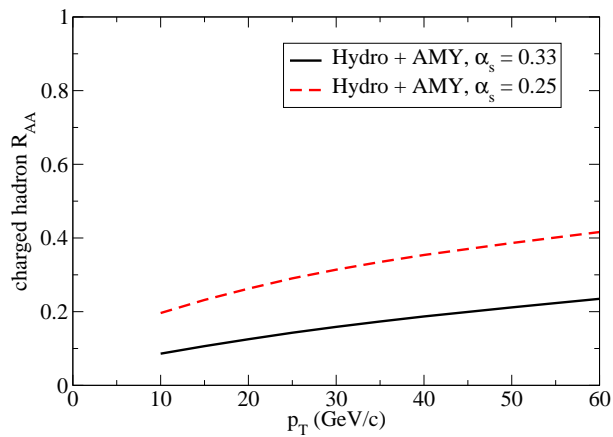}
\end{center}
\caption{\label{raa} The $p_T$ dependence of the nuclear modification factor $R_{AA}$ for charged hadrons in
central Pb+Pb collisions at mid-rapidity at the LHC.}
\end{figure}

In Fig.~\ref{raa} we present a prediction for charged hadron $R_{AA}$ as a function of $p_T$ at mid-rapidity for
central collisions at the LHC. We consider that these two values of $\alpha_s$ define a sensible band of physical
parameters.

Acknowledgments: We thank the authors of \cite{Eskola:2005ue} for providing their hydrodynamical evolution
calculation at RHIC and LHC energies, T. Renk for discussions, and the Natural Sciences and Engineering Research
Council of Canada for support.

\end{document}